\documentclass{aa}  
\usepackage{graphicx}
\usepackage{txfonts}
\begin{document}
   \title{}

   \subtitle{IGR~J18483$-$0311: an accreting X-ray pulsar observed by INTEGRAL} 

   \author{V. Sguera\inst{1}, A. B. Hill\inst{1}, A. J. Bird\inst{1}, A. J. Dean\inst{1}, A. Bazzano\inst{2}, P. Ubertini\inst{2}, N. Masetti\inst{3}, R. Landi\inst{3}, 
  A. Malizia\inst{3}, D. J. Clark\inst{1}, M. Molina\inst{1}.
          }

   \offprints{sguera@astro.soton.ac.uk}
   \institute{School of Physics and Astronomy, University of Southampton, Highfield, SO17 1BJ, UK \and 
   INAF/IASF Roma, via Fosso del Cavaliere 100, 00133 Roma, Italy \and 
   INAF/IASF Bologna, via Piero Gobetti 101, I-40129 Bologna, Italy
         }

   \date{Received 17 November 2006 / accepted 10 February 2007}


  \abstract
   {IGR~J18483$-$0311 is a poorly known transient hard X-ray source discovered by INTEGRAL during observations of the Galactic Center region 
    performed between 23--28 April 2003.}
   {To detect new outbursts from IGR~J18483$-$0311  using INTEGRAL and archival \emph{Swift} XRT observations and
    finally  to characterize the nature of this source using the optical/near$-$infrared (NIR) information available through catalogue searches.}
   {We performed an analysis of light curves and spectra of INTEGRAL and archival \emph{Swift} XRT data
    as well as of optical/NIR catalogues.}
   {We report on 5 newly discovered outbursts from IGR~J18483$-$0311 detected by INTEGRAL. 
    For two of them it was possible to constrain 
    a duration of the order of a few days. The strongest outburst reached a peak flux of $\sim$120 mCrab (20--100 keV);
    its broad band JEM--X/ISGRI spectrum (3--50 keV) is best fitted by an absorbed cutoff power law with 
    $\Gamma$=1.4$\pm$0.3, cutoff energy of 22$^{+7.5}_{-4.5}$ keV and N$_{H}$=9$^{+5}_{-4}$$\times$10$^{22}$ cm$^{-2}$. 
    Timing analysis of INTEGRAL data allowed us to identify periodicities of 18.52 days and 21.0526 seconds which are likely the orbital 
    period of the system and the spin period of the X-ray pulsar respectively.
    \emph{Swift} XRT observations of IGR~J18483$-$0311 provided a very accurate source position which strongly indicates a highly reddened star in the USNO--B1.0 and 2MASS catalogues  as its possible optical/NIR counterpart.}
   {The X-ray spectral shape, the periods of 18.52 days and 21.0526 seconds, the high intrinsic absorption, the location in 
   the direction of the Scutum spiral arm and the highly reddened optical object 
   as possible counterpart, all favour the hypothesis that IGR~J18483$-$0311 is a HMXB with a 
   neutron star as compact companion. The system is most likely a Be X-ray binary, but a Supergiant Fast X-ray Transient nature can not be entirely excluded.}


   \maketitle

\section{Introduction}
Since its launch in 2002, the INTEGRAL satellite has discovered  more than one hundred new hard X-ray sources, mainly located toward
the inner regions of the Galaxy which continue to be  extensively monitored. Many of them are characterized by absorbed hard spectra,
with little or no detectable emission in the soft X-rays since they are heavily absorbed by the interposing material. 
This, together with their often transient nature, explains why they have not been detected by any previous X-ray mission.
As discussed by Dean et al. (2006), most of the newly discovered  INTEGRAL sources should be high mass 
X-ray binaries (HMXBs), although Masetti et al. (2006a) found that several of them are actually  
Active Galactic Nuclei. 
This picture has been supported by various identifications with transient Be HMXBs or bright persistent highly absorbed supergiant HMXBs (SGXBs), 
either based on their secure identification at optical/infrared wavebands or
on their X-ray characteristics and discovery of periodic pulsations (Walter et al. 2006). As well as the highly absorbed persistent SGXBs, INTEGRAL
is also finding new members of a newly discovered class of SGXBs which escaped detection by previous X-ray missions 
mainly because of their fast X-ray transient behaviour.
They have been labeled as Supergiant  Fast X-ray Transients SFXTs (Negueruela et al. 2005, Sguera et al. 2005, Sguera et al. 2006).

\begin{table*}[t!]
\begin{center}
\caption {Summary of IBIS detections of newly discovered outbursts from IGR~J18483$-$0311.}
\label{tab:main_outbursts}
\begin{tabular}{cccccccc}
\hline
\hline
No. & Date  &  energy band & peak flux      & peak luminosity $\star$  & duration  & No. ScWs & Exp. time  \\
    &       &   (keV)           & (mCrab) &            (erg s$^{-1}$)                       &  (days)  & & (ks)  \\
\hline
1  & 19 April 2006 & 20--100 & $\sim$ 120   &  $\sim$ 7.8$\times$10$^{36}$    & $\sim$ 1.8 &  73 &  146.1 \\
2  & 5 September 2004 &  20--60 & $\sim$ 95 &  $\sim$ 4.35$\times$10$^{36}$   &   $\sim$ 0.46$\dagger$ & 21 & 35.16 \\
3  & 26 April 2004 & 20--60 & $\sim$ 80 & &                       $\sim$ 3  & 54 & 188.66\\
4  & 18 March 2004 & 20--40 & $\sim$ 135 & &                      $\sim$0.33$\dagger$  & 16 & 28.29 \\  
5  & 11 May 2003 & 20--60 & $\sim$ 75 &     $\sim$ 3.2$\times$10$^{36}$  &  $\sim$1.3$\dagger$ & 50 & 104.05 \\
\hline
\end{tabular}
\end{center}
$\dagger$ = lower limit on the duration. \\ 
$\star$  = assuming a distance of $\sim$ 5.7 kpc (see Section 5). \\
\end{table*}

The transient X-ray source IGR~J18483$-$0311 was discovered with the IBIS instrument (Ubertini et al. 2003) on board the INTEGRAL satellite 
during observations of the Galactic Center field performed between 23--28 April  2003 (Chernyakova et al. 2003).
The average X-ray flux was $\sim$10 mCrab and $\sim$5 mCrab in the energy bands 15--40 keV and 40--100 keV respectively. 
A possible X-ray outburst was observed on 26 April, when the X-ray flux increased up to $\sim$40 mCrab (15--40 keV).
Subsequently its discovery, Molkov et al (2003a) reported another detection of the source by INTEGRAL on 5 April 2003.
IGR~J18483$-$0311 has also been detected by INTEGRAL at an average flux of 4.3$\pm$0.2 mCrab (18--60 keV) 
during a survey of the Sagittarius arm tangent region in the spring 2003 (Molkov et al. 2003b).
Stephen et al. (2006) reported the association between IGR~J18483$-$0311 and the ROSAT HRI source 1RXH~J184817.3-031017.
Two optical USNO--B1.0 objects were found within the ROSAT positional uncertainty (Monet et al. 2003), one of which is also a 
near--infrared (NIR) 2MASS source (Skrutskie et al. 2006). \\
In this paper, we report in section 3.1 on IBIS detections of 5 newly discovered outbursts from IGR~J18483$-$0311,
not reported in the literature so far. Moreover, through timing analysis of the INTEGRAL light curves, we identify two 
periodicities which we interpret as being most likely attributable to the orbital period of the system (section 3.2)
and the spin period of an X-ray pulsar (section 3.3). 
We report in section 4 on two archival \emph{Swift} XRT observations of IGR~J18483$-$0311 which provide a very accurate source position. 
This allows us to  pinpoint   a highly reddened USNO-B1.0 star as the likely counterpart of the source, and its 
optical/NIR properties are discussed in section 5.
 
\section{INTEGRAL data analysis} 
IGR~J18483$-$0311 was observed by 
JEM$-$X (Lund et al. 2003) and by IBIS/ISGRI (Lebrun et al. 2003). 
The reduction and analysis of the data have been performed 
using the INTEGRAL Offline Scientific Analysis (OSA) v.5.1.
INTEGRAL observations are typically divided into short pointings (Science Windows, ScWs) 
of $\sim$ 2000 seconds duration.

We performed an analysis at the ScW level of the deconvolved ISGRI shadowgrams searching for outburst activity from IGR~J18483$-$0311.
Our ScW data set consists of all Core Program observations (the Galactic Plane Survey and the Galactic Centre Deep Exposure) from
revolution 45 (end of February 2003) to 430 (end of April 2006) as well as all public data released up to revolution 260.
The search provided 5 newly discovered outbursts from IGR~J18483$-$0311.
Due to possible cross-talk between objects in the same field of view (FOV),   
we have also investigated the variability pattern of all other bright sources in the FOV,
in addition to that of  the source of  interest. They have shown a different time variability enabling us to conclude that the light curves  obtained 
for  the  source of  interest are reliable.
Images from the X-ray monitor JEM$-$X were created for all newly discovered outbursts reported in this paper. 
Only in one case (outburst No.1 in table 1) was the source inside the JEM$-$X FOV so that it was possible to extract a spectrum and a X-ray light curve.

\section{INTEGRAL results} 
\begin{figure}[t!]
\centering
\includegraphics[width=9cm]{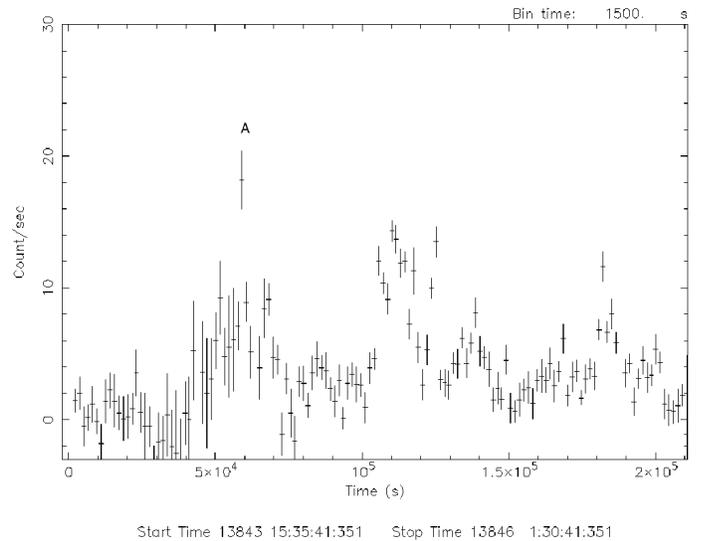}
\caption{ISGRI light curve (20--100 keV) of a newly discovered outburst of IGR~J18483$-$0311 that occurred in April 2006 (No.1 in Table 1).
The strongest flare is labeled with the letter A.}
\end{figure}
\subsection{IBIS detections of 5 newly discovered outbursts from  IGR~J18483$-$0311}
Table 1 lists the 5 newly discovered outbursts from IGR~J18483$-$0311 detected by IBIS and discussed in the present paper. 
It provides the date and energy range of the detection, the flux and luminosity at the peak, the duration of the outburst activity, the number of ScWs
during which the source is detected and the net exposure time. 

Fig. 1 shows the 20--100 keV ISGRI light curve of outburst No. 1. 
At the beginning there is no significant detection since the count rate is consistent with zero count/sec, then suddenly on 19 April 2006  $\sim$04:50 UTC
the source turned on. Initially,  the outburst activity was characterized by a prominent flare (labeled as A in Fig. 1) 
which reached  a peak flux of $\sim$ 120 mCrab or 2.04$\times$10$^{-9}$ erg cm$^{-2}$ s$^{-1}$ (20--100 keV) in a few hours and 
then dropped with the same timescale. This first flare was  followed by several others, then the source turned off. 
The duration of the total outburst activity was $\sim$1.8 days.

We combined all the ScWs during outburst No. 1 into 
a single mosaic significance map. IGR~J18483$-$0311 was detected at $\sim$50$\sigma$ 
(20--100 keV), and its coordinates (J2000) are RA=18$^{h}$ 48$^{m}$ 15.07$^{s}$, Dec=-03$^{\circ}$ 10$^{'}$ 17$^{''}$, with an error radius of 1.$^{'}$4.

The combined JEM$-$X/ISGRI spectrum (3--50 keV) from the whole duration of the outburst No.1 is best fitted by an absorbed cutoff power law 
($\chi^{2}_{\nu}$=1.19, 140 d.o.f.) with $\Gamma$=1.4$\pm$0.3,  N$_{H}$=9$^{+5}_{-4}$$\times$10$^{22}$ cm$^{-2}$ and 
cutoff energy equal to 22$^{+7.5}_{-4.5}$ keV. The absorption exceeds the galactic one along the line of sight (1.6$\times$10$^{22}$ cm$^{-2}$),
suggesting that most of the low energy absorption is intrinsic to the source.
Fig. 2 displays the absorbed cutoff power law unfolded spectrum, while  
Fig. 3 shows the contour plot for the photon index and the cutoff energy.
To account for a cross-calibration mismatch between the two instruments we have introduced a constant in the fit, 
which when left free to vary provides a value of 1.3$\pm$0.15.
It is worth pointing out that  a similarly good fit ($\chi^{2}_{\nu}$=1.2, 141 d.o.f.)  
is also provided by an absorbed bremsstrahlung model with kT=21.5$^{+2.5}_{-2}$ keV and 
N$_{H}$=7.5$^{+2.5}_{-2}$$\times$10$^{22}$ cm$^{-2}$. The latter is compatible, within the uncertainties, with the value obtained from the 
absorbed cutoff power law spectral model.

\begin{figure}
\centering
\includegraphics[width=6cm,angle=270]{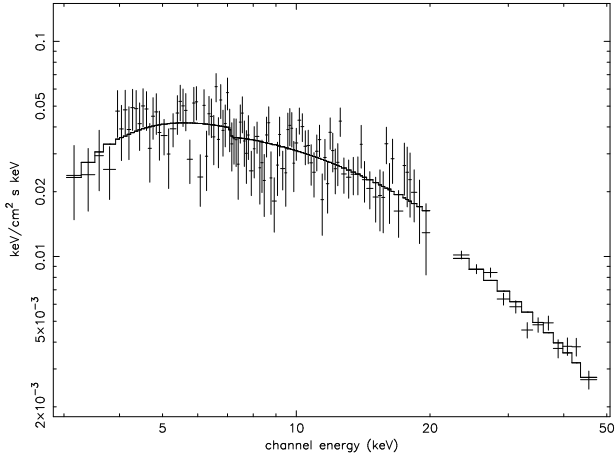}
\caption{Unfolded combined JEM$-$X and ISGRI spectrum (3--50 keV) of  IGR~J18483$-$0311 during outburst No.1 in Table 1.}
\end{figure}

\begin{figure}
\centering
\includegraphics[width=6cm,angle=270]{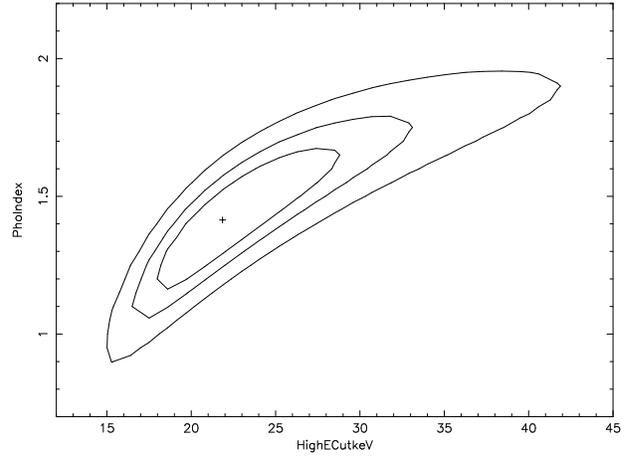}
\caption{Confidence contours level for the photon index and the cutoff energy from the spectral analysis of the outburst No.1 in Table 1.}
\end{figure}

\begin{figure}[t!]
\centering
\includegraphics[width=6cm,angle=270]{6762fig4.ps}
\caption{ISGRI light curve (20--60 keV) of a newly discovered outburst of IGR~J18483$-$0311 that occurred in September 2004 (No.2 in Table 1).}
\centering
\includegraphics[width=6cm,angle=270]{6762fig5.ps}
\caption{Unfolded bremsstrahlung spectrum (20--60 keV) of  IGR~J18483$-$0311 during outburst No.2 in Table 1.}
\end{figure}

\begin{figure}[t!]
\centering
\includegraphics[width=8cm]{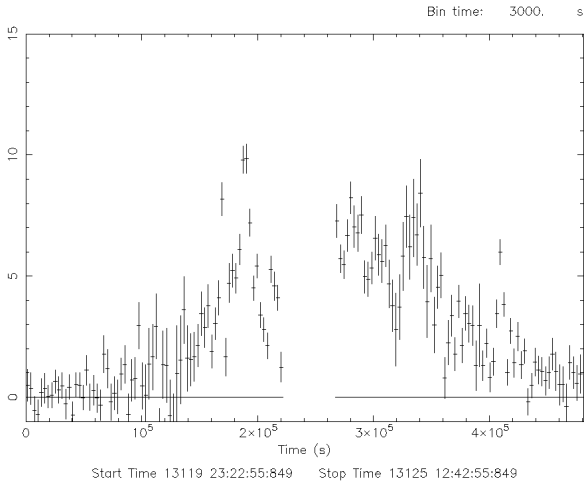}
\caption{ISGRI light curve (20--60 keV) of a newly discovered outburst of IGR~J18483$-$0311 that occurred in April 2004 (No.3 in Table 1).}
\end{figure}

Fig. 4 shows the 20--60 keV ISGRI light curve of outburst No.2 that 
started on 5 September 2004 at $\sim$  04:50 UTC and ended on the same day at  $\sim$ 16:00 UTC.
It is truncated at the beginning and at the end because the source was outside the IBIS FOV, so that it is not possible to constrain a total
duration for the outburst activity. The source was  in the IBIS FOV for $\sim$ 11 hours and its light curve is characterized by  several fast 
flares (on timescales of few tens of minutes)  reaching a  maximum peak flux  (20--60 keV) of $\sim$ 95 mCrab  or $\sim$ 1.1$\times$10$^{-9}$ erg cm$^{-2}$ s$^{-1}$.
The 20--60 keV spectrum extracted during the whole outburst is well fitted by thermal models such as black body ($\chi^{2}_{\nu}$=0.97, 14 d.o.f., 
kT=6.5$^{+0.45}_{-0.45}$ keV) or bremsstrahlung ($\chi^{2}_{\nu}$=0.61, 14 d.o.f., kT=20.6$^{+4}_{-3}$ keV).
A good fit is also provided by a simple power law with $\Gamma$=2.9$^{+0.25}_{-0.25}$ ($\chi^{2}_{\nu}$=0.74, 14 d.o.f.). Fig. 5 shows the unfolded 
20--60 keV bremsstrahlung  spectrum.

Fig. 6 displays the 20--60 keV light curve of outburst No.3  in Table 1.
The gap of $\sim$ 13 hours is due to the visibility constraint between one INTEGRAL revolution and the next.
In spite of that, the flaring activity of the source is very evident.
Initially the flux is consistent with zero, then suddenly the source turned on at $\sim$  20:00 UTC on 26 April 2004 and
reached  a maximum peak flux of $\sim$ 80 mCrab (20--60 keV); the duration of the total outburst activity was $\sim$ 3 days.
A spectrum extracted during the whole outburst (20--60 keV) is equally well fitted by a black body ($\chi^{2}_{\nu}$=0.95, 14 d.o.f., 
kT=7.2$^{+0.25}_{-0.25}$ keV) or bremsstrahlung ($\chi^{2}_{\nu}$=0.95, 14 d.o.f., kT=25.2$^{+2.5}_{-2.2}$ keV). A simple power law provides a bad 
fit ($\chi^{2}_{\nu}$=2.05, 14 d.o.f).
Fig. 7 shows the unfolded bremsstrahlung  spectrum.

The ISGRI light curve (20--40 keV) of outburst No. 4 is shown in Fig. 8.
The observations starts on 18 March 2004 at $\sim$  20:00 UTC.
The beginning  was not observed because the source was outside the IBIS FOV. Nevertheless, 
the decay of the flare is clear, in fact the 20--40 keV flux dropped from $\sim$ 135 mCrab ($\sim$ 1.02$\times$10$^{-9}$ erg cm$^{-2}$ s$^{-1}$)
to just a few mCrab in $\sim$ 4 hours. 
This strongly suggests that the source was active before entering the IBIS FOV.
Another flare is present  in the light curve (labeled as A): it reached a peak flux of $\sim$ 80 mCrab (20--40 keV) in $\sim$ 1 hour 
and then dropped  to an almost null flux with the same timescale. 
The remaining part of the light curve shows no flaring activity from the source.
The 20--60 keV spectrum extracted during its whole outburst cannot be fitted by a simple power law ($\chi^{2}_{\nu}$=1.59, 14 d.o.f).
Reasonable fits were obtained by  thermal models such as black body ($\chi^{2}_{\nu}$=0.85, 14 d.o.f., 
kT=7.8$^{+0.6}_{-0.6}$ keV) or bremsstrahlung ($\chi^{2}_{\nu}$=1.26, 14 d.o.f., kT=32.5$^{+10.5}_{-6.7}$ keV).
Fig. 9 shows the unfolded bremsstrahlung spectrum.

\begin{figure}[t!]
\centering
\includegraphics[width=6cm,angle=270]{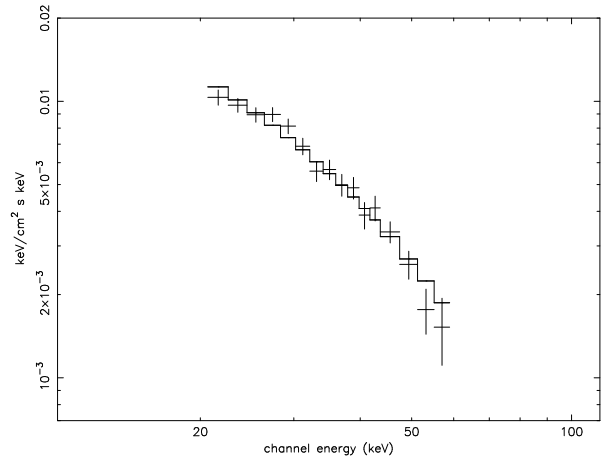}
\caption{Unfolded bremsstrahlung spectrum of outburst No.3 in Table 1.}
\end{figure}

\begin{figure}[t!]
\centering
\includegraphics[width=8cm]{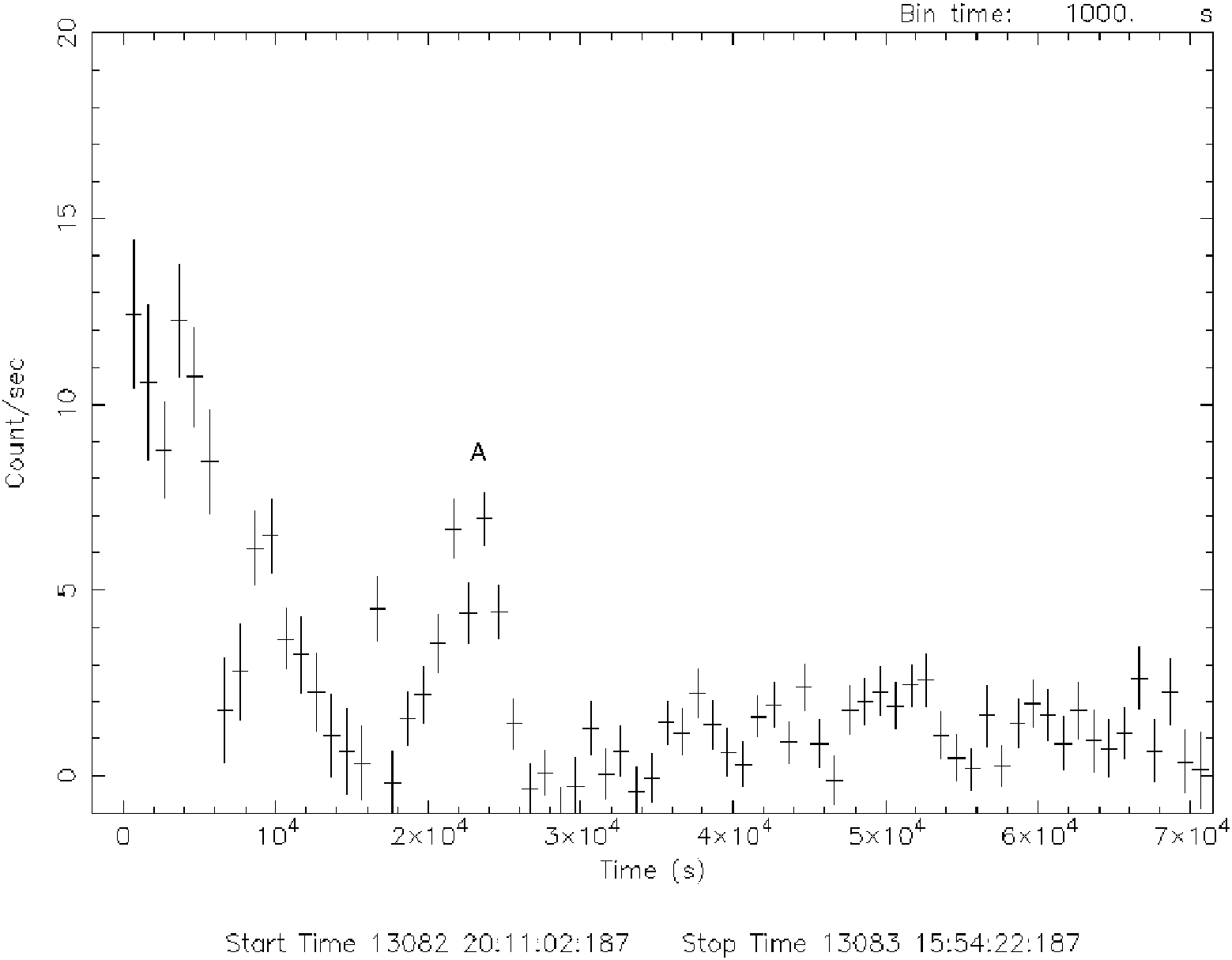}
\caption{ISGRI light curve (20--40 keV) of a newly discovered outburst of IGR~J18483$-$0311 that occurred in March 2004 (No. 4 in Table 1).}
\centering
\includegraphics[width=6cm,angle=270]{6762fig9.ps}
\caption{Unfolded bremsstrahlung spectrum of outburst No.4 in Table 1}
\end{figure}

\begin{figure}[t!]
\centering
\includegraphics[width=8cm]{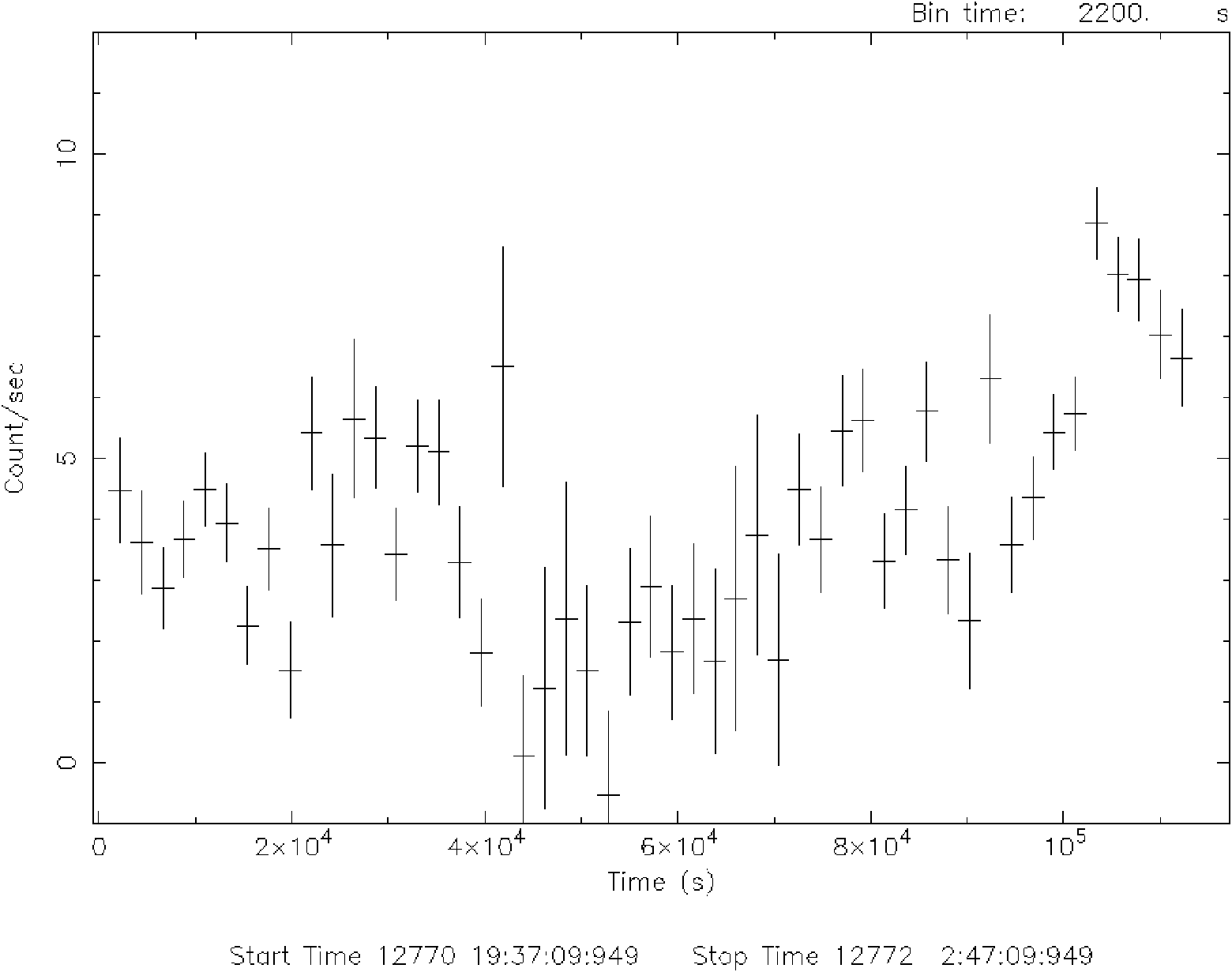}
\caption{ISGRI light curve (20--60 keV) of a newly discovered outburst of IGR~J18483$-$0311 that occurred in May 2003 (No.5 in Table 1).}
\end{figure}
\begin{figure}[t!]
\centering
\includegraphics[width=6cm,angle=270]{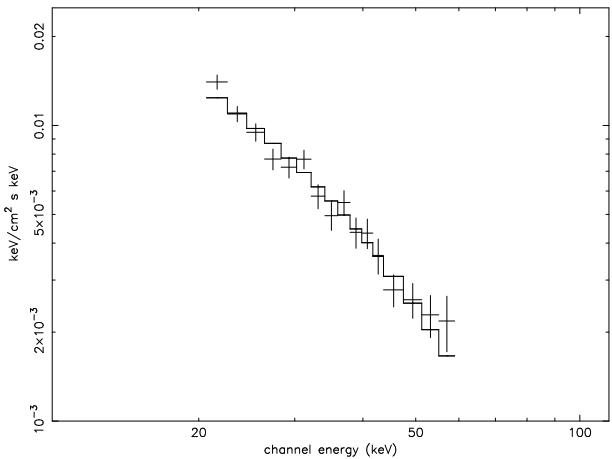}
\caption{Unfolded bremsstrahlung spectrum (20--60 keV) of outburst No.5 in Table 1 of IGR~J18483$-$0311.}
\end{figure}

\begin{table*}[t!]
\begin{center}
\caption {Summary of spectral analysis of the 5 newly discovered outbursts from IGR~J18483$-$0311. The name of the spectral models are 
written in XSPEC terminology.}
\label{tab:spectral_outbursts}
\begin{tabular}{ccccccc}
\hline
\hline
Model        & parameter                      &  outburst No.1  & outburst No.2   &outburst No.3 &  outburst No.4 & outburst No.5    \\
             &                                &  (3--50 keV)    & (20--60 keV)    &  (20--60 keV)&   (20--60 keV) &  (20--60 keV)      \\
\hline
wa+cutoffpl  & N$_{H}$(10$^{22}$ cm$^{-2}$)   &  9$^{+5}_{-4}$  &                 &              &                &                \\
             & $\Gamma$                       &  1.4$\pm$0.3    &                 &              &                &                \\
             & E$_{c}$                        &  22$^{+7.5}_{-4.5}$               &                 &              &                &                \\  
             &  $\chi^{2}_{\nu}$(d.o.f.)      &  1.19 (140)     &                 &              &                &                \\  
\hline
wa+bremss    & N$_{H}$(10$^{22}$ cm$^{-2}$)   &  7.5$^{+2.5}_{-2}$  &                 &              &                &                \\
             & kT                             &  21.5$^{+2.5}_{-2}$ &                 &              &                &                \\
             &  $\chi^{2}_{\nu}$(d.o.f.)      &  1.2 (141)          &                 &              &                &                \\  
\hline
bremss       &  kT                            &  26$^{+3.3}_{-2.5}$   & 20.6$^{+4}_{-3}$ &  25.2$^{+2.5}_{-2.2}$  & 32.5$^{+10.5}_{-6.7}$ & 22$^{+2}_{-2.5}$      \\ 
             &  $\chi^{2}_{\nu}$(d.o.f.)      & 1.49 (142)            & 0.61 (14)        &  0.95 (14)             & 1.26 (14)             & 1.1 (14)              \\
\hline
bb           &  kT                            &                     & 6.5$^{+0.45}_{-0.45}$  & 7.2$^{+0.25}_{-0.25}$  & 7.8$^{+0.6}_{-0.6}$ &                \\
             &  $\chi^{2}_{\nu}$(d.o.f.)      &  4.86 (142)         & 0.97 (14)              & 0.95 (14)              & 0.85 (14)           & 3.23 (14)       \\
\hline
po           & $\Gamma$                       &           & 2.9$^{+0.25}_{-0.25}$  &                       &                         & 2.9$^{+0.15}_{-0.15}$       \\
             &  $\chi^{2}_{\nu}$(d.o.f.)      & 2.65 (142)& 0.74 (14)              & 2.05 (14)             & 1.59 (14)               & 0.85 (14)               \\
\hline
\hline
\end{tabular}
\end{center}
\end{table*} 

Finally, Fig. 10 shows the 20--60 keV ISGRI light curve of outburst No.5 in Table 1.
It started on 11 May 2003 at $\sim$  19:30 UTC and it ended
on 13 May 2003 at $\sim$  02:30 UTC.
It is truncated at the beginning and at the end because the source was outside the IBIS FOV, so it is not possible to constrain a total 
duration of the outburst. The peak flux was  $\sim$ 75 mCrab (20--60 keV).
The 20--60 keV spectrum of the whole  outburst is best fitted by a bremsstrahlung model ($\chi^{2}_{\nu}$=1.1, 14 d.o.f)
with  kT=22$^{+2}_{-2.5}$ keV, whereas a black body is a very bad description to the data ($\chi^{2}_{\nu}$=3.2, 14 d.o.f).
A reasonable fit is also provided by a simple power law ($\chi^{2}_{\nu}$=0.85, 14 d.o.f) with $\Gamma$= 2.9$^{+0.15}_{-0.15}$. 
Fig. 11 shows the unfolded 20--60 keV bremsstrahlung spectrum.

Table 2 is a summary of the spectral analysis of each newly discovered outburst.
Outburst No.1 is the only one for which a broad band spectrum (3--50 keV) is available. The cutoff power law model constrains very well its spectral properties 
both at soft and hard X-rays, with best fit parameters  $\Gamma$=1.4$\pm$0.3, E$_{c}$=22$^{+7.5}_{-4.5}$ keV
and N$_{H}$=9$^{+5}_{-4}$$\times$10$^{22}$ cm$^{-2}$.
Moreover, it can be noted that a bremsstrahlung model provided a good description of all 5 outbursts with a similar temperature kT in the range 20--32 keV.
A black body  gave good fits for the outbursts No. 2, 3 and 4 with a similar temperature kT in the range 6.5--7.5 keV, 
but is unacceptable in the case of the outbursts No. 1 and 5.
Finally, a simple power law described well the spectra of outburst No. 2 and 5 providing the same value of the photon index ($\sim$2.9)
Whereas, outbursts No. 3 and 4  could not be described by a simple power law.

\subsection{Recurrence timescale}
\begin{figure}[t!]
  \centering 
  \includegraphics[width=0.95\linewidth,clip]{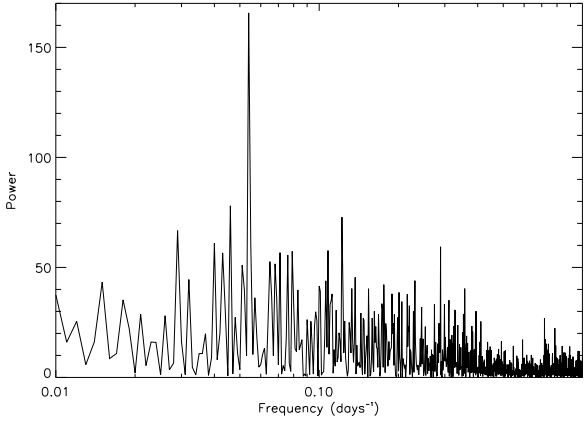}
  \caption{Lomb-Scargle periodogram generated from the 20--40 keV light
  curve of IGR J18483$-$0311}
  \label{LSpow}
\end{figure}

We investigated the long term light curve of IGR J18483$-$0311 searching for any evidence of periodicity
which would indicate the recurrence timescale of its  transient
activity. The flux of IGR J18483$-$0311 was extracted from each ISGRI pointing
where the source was within 12$^\circ$ of the centre of the field of
view, producing a long term light curve of the source on the ScW
timescale. 
A 12$^\circ$ limit was applied because the off-axis response of ISGRI is not well modelled at large off-axis angles and consequently 
in combination with the telescope dithering (or the movement of the source within the FOV) 
introduces a systematic error in the measurement of source fluxes generating
problems in the detection of periodic signals (Hill, 2006). The 20--40 keV light curve was then searched
for periodicities using the Lomb-Scargle periodogram method by means
of the fast implementation of Press \& Rybicki (1989) and Scargle (1982).
The resulting power spectrum is shown in Fig.~\ref{LSpow}; the peak
power of 165.7 corresponds to a frequency of 0.0540 days$^{-1}$. 
This frequency equates to a period of 18.52 days with a
theoretical Lomb-Scargle error of $\pm$ 0.01 days.
The error on the angular frequency is given by Horne \& Baliunas (1986).

\begin{equation}\label{equ:freq_err}
	\delta\omega = \frac{3 \pi \sigma_{N}}{2 \sqrt N T A}
\end{equation}

where $\sigma$$_{N}$$^{2}$ is the variance of the noise, N is the
number of data points, T is the total length of the data set and A is
the amplitude of the signal given by:

\begin{equation}\label{equ:amp}
	A = 2 \sqrt{\frac{z_{0} \sigma_{s}^{2}}{N}}
\end{equation}

\begin{figure}[t!]
  \centering 
  \includegraphics[width=0.95\linewidth, clip]{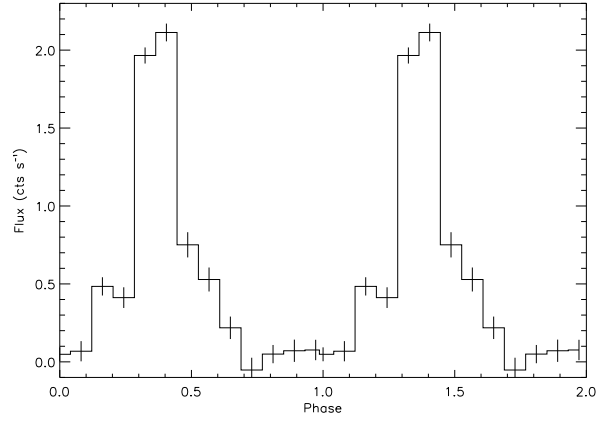}
  \caption{The phase-folded background subtracted 20--40 keV light curve of IGR J18483$-$0311. The data are folded on a period of 18.52 days.}
  \label{fold}
\end{figure}

 \begin{table*}[tbhp]
   \begin{center}
     \caption{Analysis of the 20--40 keV light curve of IGR~J18483$-$0311 around the expected
     start time of outbursts based upon the 18.52 days periodicity. For the
     outburst numbers not listed there was no significant coverage of the source by
     INTEGRAL around those times.}
    \begin{tabular}[h]{llll} \hline
      Outburst No. & Expected start time (MJD) &   INTEGRAL coverage (MJD) & Comments\\
      \hline \hline
      1 & $\sim$ 52714.48 & 52714.99-- 52715.63 &  Low level of outburst activity (maximum peak flux $\sim$ 20 mCrab)\\
      2 & $\sim$ 52733    & 52734.92-- 52735.61 & Outburst reported by Molkov et al. 2003a (peak flux $\sim$ 40 mCrab) \\
      3 & $\sim$ 52751.52 & 52753.04-- 52757.50 &  Outburst when the source was discovered (Chernyakova et al. 2003) \\ 
      4 & $\sim$ 52770.04 & 52770.80-- 52772.08 & Outburst discussed in this paper (No.5 in table 1)\\
      21 & $\sim$ 53084.88 & 53082.85-- 53083.62 & Outburst discussed in this paper (No.4 in table 1)\\
      23 & $\sim$ 53121.92 & $\sim$ 53121.7$\star$ & Outburst discussed in this paper (No.3 in table 1)\\
      30 & $\sim$ 53251.56 & 53253.20--53253.60 & Outburst discussed in this paper (No.2 in table 1)      \\ 
      62 &                 &  $\sim$ 53844.2$\star$ & Outburst discussed in this paper (No.1 in table 1) \\
      \hline
      \end{tabular}
     \label{tab:burst_fit}
     \end{center}
$\star$ = start time of the outburst 
   \end{table*}

where z$_{0}$ is the Lomb-Scargle power and $\sigma$$_{s}$$^{2}$ is
the variance of the light curve. The period estimate
was confirmed with a Monte-Carlo simulation.  Each flux measurement was
adjusted using Gaussian statistics within its individual error
estimate to generate a simulated light curve of the source.  The
corresponding periodogram was produced and any detected periodicity
recorded.  We simulated 200,000 light curves in this fashion and found
that 99.5\% identified a periodicity centred at 18.52 days.
The phase-folded 20--40 keV lightcurve of IGR J18483$-$0311 is shown in
Fig.~\ref{fold}.  A clear flare-like profile is apparent, with the
source being predominantly undetected and then briefly flaring to a
detectable level.  The average duration of an outburst can be seen to be
of the order of 3 days (FWHM). This is consisted with the measured durations for outburst No. 1 and 3 in Table 1.

\subsubsection{Looking for missed outbursts}

The detected periodicity of $\sim$ 18.52 days indicates the recurrence
timescale of outbursts from IGR J18483$-$0311, so that 
we could search for additional weaker outbursts in the INTEGRAL
observations.  Between the first and last observation analysed in this
paper (52704 -- 53846 MJD) 62 outbursts may have occurred.  Using the
measured start time (MJD$\sim$53844.2) of outburst No. 1 in Table 1 and the periodicity of 18.52
days, the full long term light curve of IGR J18483$-$0311 was examined around 
the predicted outburst start times. To this aim, no  off--axis angle limit was applied during the extraction of full long term ISGRI light curve.
The results of this examination are  shown in Table~\ref{tab:burst_fit}.  
Of the 62 potential outbursts, only  8 were well covered by INTEGRAL observations. For the outburst numbers not listed there was no significant coverage of the source
by INTEGRAL around those times. These 8 outbursts include the 5 reported brightest ones in Table 1
and discussed in this paper, 2 outbursts with a low level of activity and the outburst when the source was discovered by Chernyakova et al. (2003).
It should be noted that all but one of the outbursts listed in Table 2 occured when expected as based upon the 18.52 days periodicity.
The only exception (No. 21 in Table 2 or No.4 in Table 1) occured at least 2 days earlier than
expected. This may in part be explained by the difficulty in identifying a period when each outburst has a unique shape, moreover 
only 2 out of 5 outbursts detected by IBIS have been seen for their entire duration. Consequently 
the statistical error assigned to the 18.52 days period is likely to be underestimate.  
However, it might also indicate that the outbursting behaviour of the source is only
semi-regular, implying  that there may be something which modulates the outburst time scale beyond the mechanism
that produces the periodicity.  This behaviour could be similar to that observed in Be X-ray transients in which the outbursts are
associated with both the orbital period of the system and with the size of the disk around the Be star.

\subsection{Searching for a pulse period}
\begin{figure}[t!]
  \centering 
  \includegraphics[width=0.95\linewidth,clip]{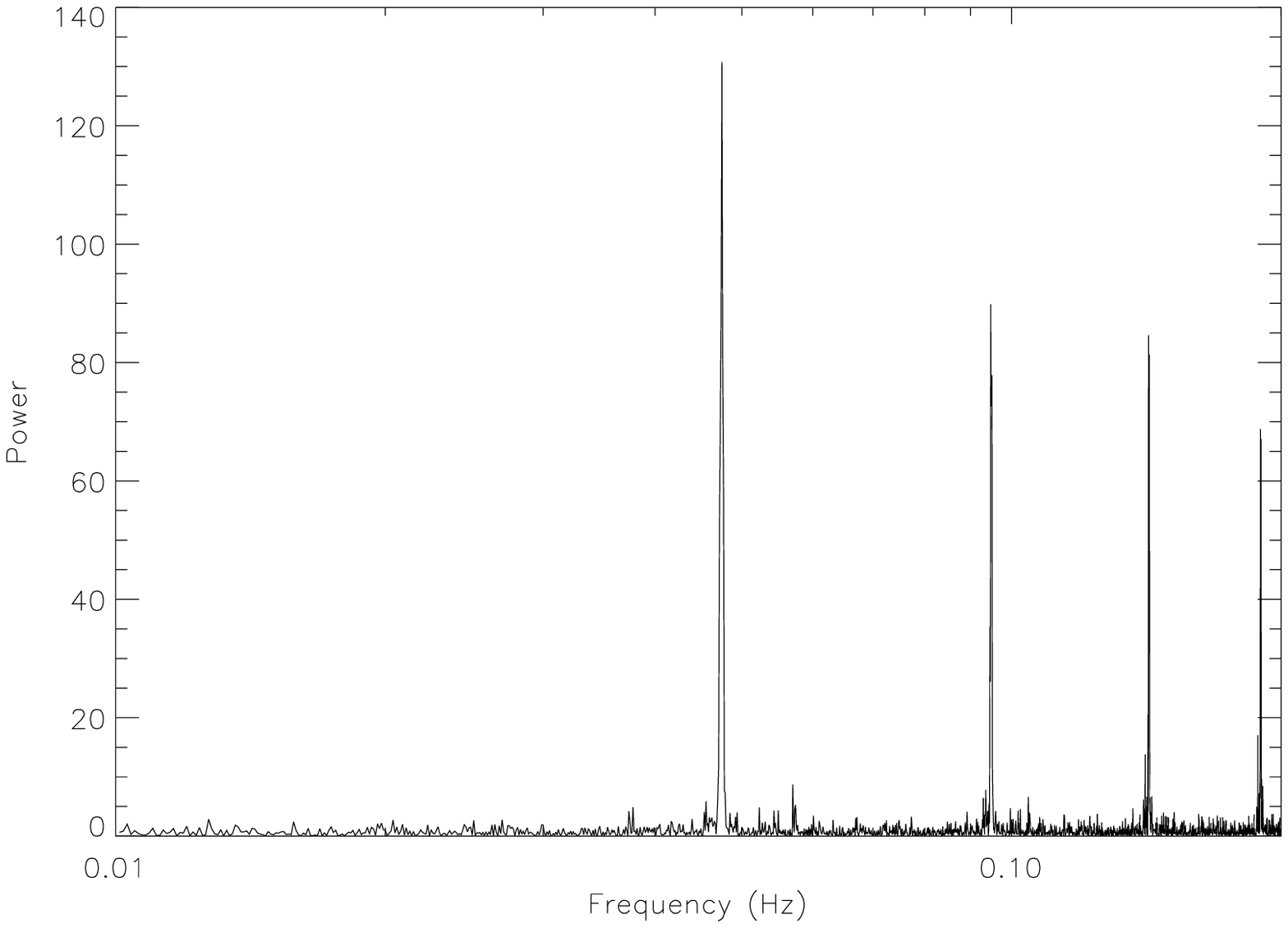}
  \caption{Lomb-Scargle periodogram generated from the JEM-X 4--20 keV light
  curve of IGR J18483$-$0311 during the outburst No.1 in Table 1. The periodicity at 0.04750 Hz is clearly apparent as are the 1$^{st}$, 2$^{nd}$ and 3$^{rd}$ harmonics.}
  \label{fig:spin_psd}

  \centering 
  \includegraphics[width=0.95\linewidth, clip]{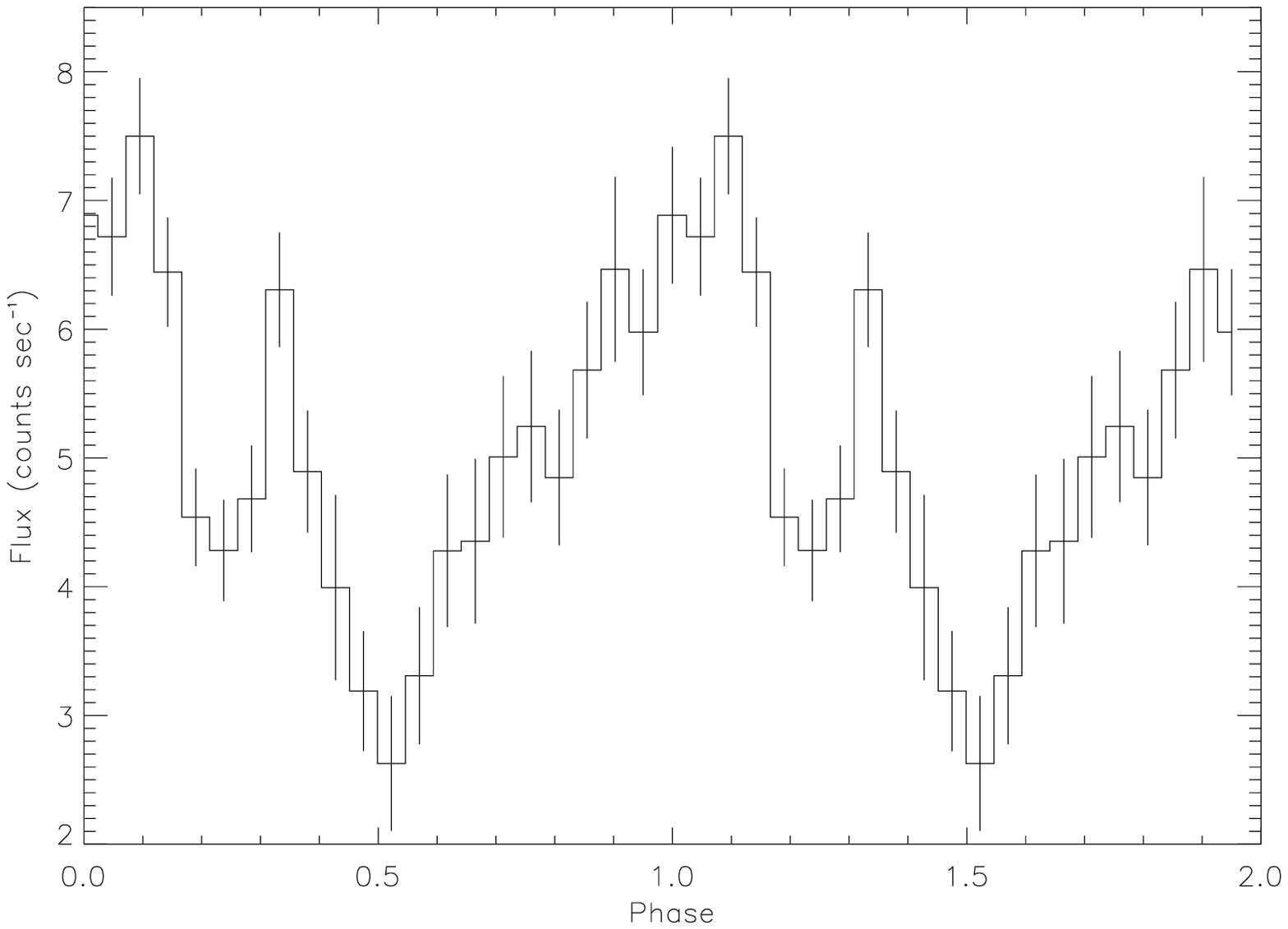}
  \caption{The phase-folded 4--20 keV light curve of IGR J18483$-$0311. The data are folded on the detected  period of 21.0526 seconds.}
  \label{fig:spin_fold}
\end{figure}
Fine timing resolution ISGRI light curves (20--40 keV) were constructed for the duration  of the two brightest outbursts (No. 1 and No. 4)
in Table~\ref{tab:main_outbursts}, using the \emph{ii--light} tool in OSA5.1.
A search for pulsations was performed using the 
Lomb-Scargle periodogram method with no positive results.

However, IGR J18483$-$0311 was also inside the JEM-X FOV during part of the outburst No.1 in Table 1.
Consequently, we produced a 
light curve in the 4--20 keV band using the standard OSA5.1 tools. 
This light curve was also searched for periodicities  using the Lomb-Scargle method; the resulting power spectrum is shown in Figure~\ref{fig:spin_psd}.  
The peak power of 130.8 corresponds to a frequency of 0.04750 Hz.  The error on this measurement is calculated using 
equations~\ref{equ:freq_err} \&~\ref{equ:amp}.  The corresponding period is 21.0526 $\pm$ 0.0005 seconds; 
the 1$^{st}$, 2$^{nd}$ and 3$^{rd}$ harmonics of this period are also evident in the power spectrum. 

Folding the  JEM-X light curve on the 21.0526 seconds yields the pulse-period phase-folded light curve shown in Figure~\ref{fig:spin_fold}.  
We attribute this periodicity to the spin period of an X-ray pulsar located within the system. 
There are no significantly detected features in the profile.  
The pulse fraction, $(I_{max} - I_{min})/I_{max}$, where $I_{max}$ and $I_{min}$ are the maximum 
and minimum count rates respectively, is $\sim$ (65$\pm$10) \%.

\section{\emph{Swift} observations and results} 

\begin{table*}[t!] 
\begin{center}
\caption {Summary of \emph{Swift} XRT  observations of IGR~J18483$-$0311.}
\begin{tabular}{cccccc}
\hline
\hline
 Obs date  & closest expected outburst  & average absorbed flux &  average luminosity$\star$  & $\Gamma$ & $N_{H}$  \\
 (MJD)     & (MJD) &  (erg cm$^{-2}$ s$^{-1}$) &  (erg s$^{-1}$)            &          &  (cm$^{-2}$)  \\
\hline
  53782 & $\sim$ 53788.64   &  $\sim$ 4.3$\times$10$^{-12}$ (1--7 keV)&  $\sim$ 1.67$\times$10$^{34}$  & 1.7$\pm$0.7    &    6.3$^{+2.4}_{-1.8}$$\times$10$^{22}$    \\
   53799 & $\sim$ 53807.16  &  $\sim$ 2.2$\times$10$^{-11}$ (1--7 keV)&  $\sim$ 8.55$\times$10$^{34}$       & 1.3$\pm$0.25   &    4.8$\pm$0.9$\times$10$^{22}$ \\
\hline
\end{tabular}
\end{center}
$\star$  = assuming a distance of $\sim$ 5.7 kpc (see Section 5) \\
\end{table*} 

In this section we report on  X-ray observations acquired with the XRT 
(X-ray Telescope) on board the 
\emph{Swift} satellite (Gehrels et al. 2004). A search of the XRT data archive revealed that \emph{Swift} carried out  2 
observations of IGR~J18483$-$0311, on 16 February 2006 and 5 March 2006.
The XRT collected data for a total exposure 
time of 8 ks and 5.6 ks, respectively. 
Unfortunately in both observations the source was outside the IBIS  FOV. 
XRT data reduction was performed using XRTDAS v. 2.4 standard data 
pipeline package ({\sc xrtpipeline} v. 0.10.3), in order to produce 
screened event files. All data are extracted only in the Photon Counting 
(PC) mode, 
adopting the standard grade filtering (0--12 
for PC) according to the XRT nomenclature. Events for spectral analysis 
were extracted within a circular region of radius 20$^{\prime \prime}$, 
which encloses about 90\% of the PSF at 1.5 keV (Moretti et al. 2004), 
centered on the source position.
The background was extracted from various source-free regions close to the 
X-ray source of interest using both circular/annular regions with 
different radii, in order to ensure an evenly sampled background.

First of all, the \emph{Swift} XRT  analysis of IGR~J18483$-$0311 provided a very accurate position 
(RA=18$^{h}$ 48$^{m}$ 17.17$^{s}$, Dec=-03$^{\circ}$ 10$^{'}$ 15.54$^{''}$, J2000) with 
the error radius being equal to 3$^{''}$.3. 
This refined source position was obtained  using the method reported in Moretti et al. (2006), which
allows positional errors accurate to 3$^{''}$.

For the observation of 16 February 2006 (OBS1),
it was possible to extract a meaningful spectrum only in the energy range 
1--7 keV. Above 7 keV and below 1 keV the statistics
were not good enough to perform a spectral analysis. The spectrum from 1--7 keV is best fitted by an absorbed power law
($\chi^{2}_{\nu}$=1.38, 38 d.o.f) with  $\Gamma$=1.7$\pm$0.7 and N$_{H}$=6.3$^{+2.4}_{-1.8}$$\times$10$^{22}$ cm$^{-2}$.
Thermal models such as black body or bremsstrahlung provide very bad fits with a $\chi^{2}_{\nu}$ greater than 2.
The spectrum (1--10 keV) extracted during the observation of 5 March 2006 (OBS2) 
is best fitted by an absorbed power law ($\chi^{2}_{\nu}$=1.009, 67 d.o.f) with  $\Gamma$=1.3$\pm$0.25 and N$_{H}$=4.8$\pm$0.9$\times$10$^{22}$ cm$^{-2}$.
Thermal models such as  black body or bremsstrahlung are  again a bad description of the data.
Table 4 provides a list of the characteristics of the two \emph{Swift} XRT observations of IGR~J18483$-$0311.
In order to check for variability in the spectral index between the two  \emph{Swift} observations, both absorbed power law spectra have been fitted fixing the
N$_{H}$ to the average value found from the previous spectral analysis (N$_{H}$=5.55$\times$10$^{22}$ cm$^{-2}$).
By doing so, no variability  has been found since the photon index assumed  an identical value of 1.5$\pm$0.25 (OBS1) and  1.5$\pm$0.13 (OBS2), respectively.

Taking into account the \emph{Swift} XRT  spectra of IGR~J18483$-$0311, we extrapolated its flux in the ROSAT energy range 0.1--2 keV.
A count rate of $\sim$0.00046 count/sec  was obtained, which is $\sim$ 30 times smaller than that of the ROSAT HRI source (0.0139 count/sec) associated with
IGR~J18483$-$0311 according to Stephen et al. (2006). This, together with the huge absorption of IGR~J18483$-$0311 as measured with INTEGRAL and 
\emph{Swift} XRT, casts some doubt on a possible association between the ROSAT HRI source and IGR~J18483$-$0311.

Fig. 16 shows the 0.2--10 keV light curve of the two  \emph{Swift} XRT observations, OBS1 and OBS2 (bin time 200 seconds).
The time axis is in seconds, counted from the beginning of OBS1, and the  flux axis is in count/sec.
We can note a variability since the source during OBS2 is slightly brighter than OBS1. In fact,  
the average absorbed fluxes during OBS2 and OBS1 are 2.2$\times$10$^{-11}$ (1--7 keV) and  4.3$\times$10$^{-12}$ (1--7 keV), respectively. 
It seems that in both periods the source was not in outburst but in a low state. 
Fine time resolution light curves of both \emph{Swift} XRT observations were searched for pulsations 
using the Lomb-Scargle method but  no positive results were obtained; it is likely  the statistics were not good enough because of 
the small exposure time of the observations and the low state of the source.

It is worth pointing out that the two \emph{Swift} observations occured several days before the closest expected outburst of IGR~J18483$-$0311, 
based upon the 18.52 days periodicity (see Table 4). Assuming that the \emph{Swift}  source is associated with IGR~J18483$-$0311, 
the \emph{Swift} XRT detections probably represent its quiescence emission.

However, we cannot entirely exclude that the \emph{Swift} source 
is not associated with IGR~J18483$-$0311 because of the lack of a simultaneous  \emph{Swift}--INTEGRAL detection during an outburst.

\begin{figure}[t!]
\centering
\includegraphics[width=8cm]{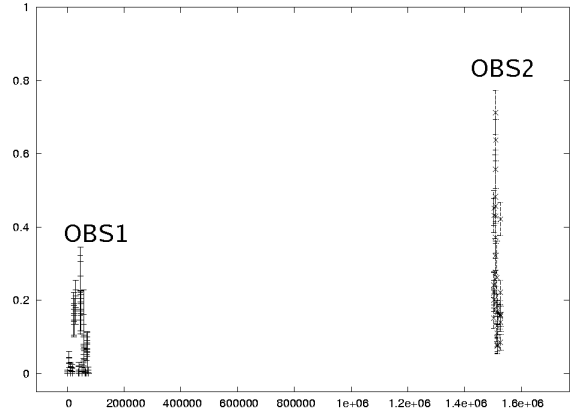}
\caption{\emph{Swift} XRT light curve (0.2--10 keV) light curve of the two  \emph{Swift} XRT observations. The bin time is 200 seconds.
Time axis is in seconds, counted from the beginning of the first \emph{Swift} XRT observation. Flux axis is in count/sec.}
\end{figure}

\section{An highly redenned star as optical counterpart of  IGR~J18483$-$0311} 
As we can note in Fig. 17, the \emph{Swift} XRT very accurate position of  IGR~J18483$-$0311 (3$^{''}$.3 error radius)
allows us to identify an optical USNO--B1.0 source as the  likely optical counterpart.
The position of this star (RA=18$^{h}$ 48$^{m}$ 17.2$^{s}$, Dec=-03$^{\circ}$ 10$^{'}$ 16.5$^{''}$, (J2000))
is 1$^{''}$.06 from the \emph{Swift} XRT location of  IGR~J18483$-$0311.
The optical and NIR magnitudes  
extracted from the USNO-B1.0  and 2MASS  catalogues, respectively, are $R$ = 19.26, $I$ = 15.32, 
$J$ = 10.74, $H$  = 9.29 and $K$ = 8.46. This indicates an extremely 
reddened object, and the optical/NIR color indices of this source  
strongly resemble those of a heavily absorbed early-type star,
similar to the case of 2RXP J130159.6$-$635806 = IGR J13020$-$6359
(Chernyakova et al. 2005), identified as a HMXB by Masetti et al.
(2006). Indeed, assuming the Milky Way extinction law (Cardelli et al. 1989),
we find that the optical/NIR color indices are consistent 
with those of a late O / early B-type star (Wegner 1994) suffering 
from a reddening of $A_V \approx$ 13 mag. This, using the formula of 
Predehl \& Schmitt (1995), implies a column density of N$_{H}$ 
$\sim$ 2.3$\times$10$^{22}$ cm$^{-2}$, which is lower than that 
inferred with the {\it Swift} and {\it INTEGRAL} spectral analysis 
results. This may indicate that part of the absorption detected in 
X--rays is local to the accreting object.
Admittedly, if one considers the $R-I$ color index alone, one gets 
a higher $V$-band optical extinction ($A_V \approx$ 16 mag).
However, it should be noted that, for faint ($>$18) magnitudes,
the photometry of USNO catalogues can have uncertainties as high as 
1 mag or more (e.g. Masetti et al. 2003). Thus, we consider the $A_V$ 
estimate obtained using the $IJHK$ magnitudes as a more correct one.
Assuming then $A_V \approx$13 mag along the IGR J18483$-$0311 line of 
sight and a B0 spectral type for the companion star in this system, 
we can infer its distance, depending on the luminosity class (main 
sequence, giant or supergiant) of the star. We find that, for these 
three cases, a distance of $\sim$2.1, $\sim$3.5 and $\sim$5.7 kpc, 
respectively, is found. The first (main sequence) case would place 
the system in the Sagittarius arm of the Galaxy (e.g. Leitch \& 
Vasisht 1998): this distance is however too close to justify the 
inferred optical absorption.

We consider it more likely that the secondary star is an early 
giant or supergiant: in this case the source would lie either in 
the near or in the far side of the Scutum-Crux arm, respectively.
However, only optical/NIR spectroscopy of this object can
help us to shed more light on its nature.

\begin{figure}[t!]
\centering
\includegraphics[width=9cm,height=5.5cm]{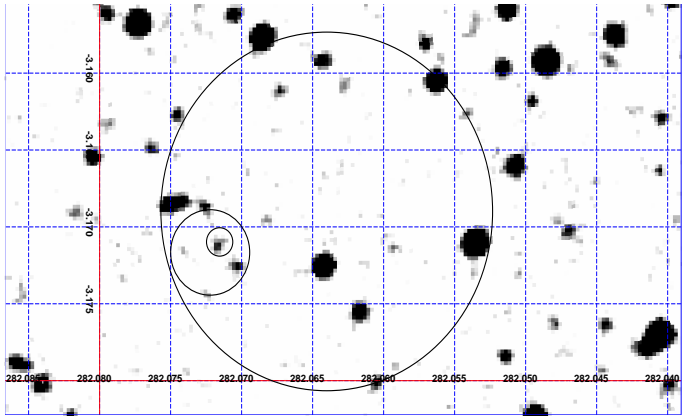}
\caption{USNO B1.0  optical field image (R2 magnitude).
The circles represent, from larger to smaller, the ISGRI error circle of IGR~J18483$-$0311 (Bird et al. 2006), ROSAT (Stephen et al. 2006) and  \emph{Swift} (this paper).
As we can note, the accurate  \emph{Swift}
position allow us to identify the likely optical counterpart for IGR~J18483$-$0311.}
\end{figure}

\section{Discussions and Conclusions} 
We report on 5 newly discovered outbursts from IGR~J18483$-$0311 detected by IBIS. 
In each case the outburst activity lasts no longer that a few days and it 
is characterized by several fast flares with timescales of a few hours. 
The broad band (3--50 keV) JEM$-$X/ISGRI spectrum of outburst No.1 is best fitted by an absorbed cutoff power law 
with  $\Gamma$=1.4$\pm$0.3, E$_{c}$=22$^{+7.5}_{-4.5}$ keV and N$_{H}$=9$^{+5}_{-4}$$\times$10$^{22}$ cm$^{-2}$. 
The latter exceeds the galactic absorption along the line of sight (1.6$\times$10$^{22}$ cm$^{-2}$)
suggesting that most of the low energy absorption is intrinsic to the source. 

Timing analysis performed on the long term ISGRI light curve allows us to identify a periodicity of 18.52 $\pm$ 0.01 days 
which is most likely a measurement of the orbital period of the system.  
Analysis of the 4--20 keV JEM-X light curve of the brightest complete outburst (No.1 in Table 1)
identifies a periodicity of 21.0526 $\pm$ 0.0005 seconds which could be due to to the spin period of an X-ray pulsar.

From \emph{Swift} XRT data analysis, we obtain a very accurate source position which finally allows 
us to pinpoint a single optical object as the probable  counterpart of IGR~J18483$-$0311. Its optical/NIR magnitudes indicate it as  
an extremely reddened object. In particular the optical/NIR color indices  
strongly resemble those of a heavily absorbed early-type star.

The X-ray spectral shape, the periodicities of 18.52 days and 21.0526 seconds, the high intrinsic absorption, the location in 
the direction of the Scutum spiral arm and the   highly reddened optical object 
as possible counterpart, support the hypothesis that IGR~J18483$-$0311 is a HMXB with a 
neutron star as compact companion. Assuming the 18.52 days periodicity as 
the system orbital period and the 21.0526 seconds periodicity as the neutron star spin period, then 
the source lies in the Be HMXB transients region of the Corbet diagram (Corbet 1986). Moreover, the characteristics of the outbursts 
and their regular recurrence at $\sim$ 18.52 days
strongly support a Be/X-ray transient HMXB nature. However, we can not entirely exclude 
the source to be a supergiant  fast X-ray transient because of the timing and spectral X-ray properties resembling those of already known 
SFXTs (Sguera et al. 2005, Sguera et al. 2006, Sguera et al. 2007). The typical duration of the outburst activity of  IGR~J18483$-$0311  is a few days, 
somewhat longer than typical outbursts from SFXTs which are usually shorter than a day, 
typically a few hours (Sguera et al. 2005, Sguera et al. 2006). However, outbursts from SFXTs lasting a few days have been 
already reported (Sidoli et al. 2006).
It is worth bearing in mind  that SFXTs are characterized by typical luminosity ratios L$_{Max}$ /  L$_{Min}$ $\sim$ 10$^{4}$.
Assuming that the \emph{Swift}  XRT detection is associated with IGR~J18483$-$0311 and 
it represents its quiescence emission, then its luminosity ratio is L$_{Max}$ /  L$_{Min}$ $\sim$ 10$^{3}$, an order of magnitude smaller than that typical of SFXTs.

Spectroscopy of the optical/NIR counterpart of IGR~J18483$-$0311 is thus essential to fully characterize 
this hard X-ray emitting object.

\begin{acknowledgements}
We thank the anonymous referee for very useful comments which helped us to improve
the paper. This research has been supported by University of Southampton School of Physics
and Astronomy. The italian co-authors acknowledge support via contract ASI/INAF I/023/05

\end{acknowledgements}

\end{document}